\documentclass{llncs}

\usepackage{amsthm,amsmath,amssymb}
\usepackage{a4}
\usepackage{url}
\usepackage{floatflt}
\usepackage{paralist}
\usepackage{ifthen}
\usepackage{graphicx}
\usepackage{subfigure}
\usepackage{wrapfig}
\usepackage{xspace}
\usepackage[usenames]{color}
\usepackage{caption}

\renewcommand{\subparagraph}{}
\usepackage[small,compact]{titlesec}

\renewcommand{\subparagraph}{}

\title{Significant Diagnostic Counterexamples in Probabilistic Model Checking}
\author{Miguel E. Andr\'es$^1$
\thanks{Supported by NWO project 612.000.526}, Pedro D'Argenio$^2$
\thanks{Supported by the ANPCyT project PICT 26135 and CONICET project PIP 6391}
, Peter van Rossum$^1$}
\institute{$^1$Institute for Computing and Information Sciences,
The Netherlands. \email{\{mandres,petervr\}@cs.ru.nl}\\
$^2$FaMAF, Universidad Nacional de C\'ordoba, CONICET, Argentina.
\email{dargenio@famaf.unc.edu.ar}}

\newcommand{\Nat}{{\mathbb N}}
\newcommand{\Var}{{\cal V}}
\newcommand{\M}{{\cal P}}

\newcommand{\F}[1]{\lozenge {#1}}
\newcommand{\G}[1]{\square {#1}}
\newcommand{\Until}[2]{{#1}{\cal U}{#2}}

\newcommand{\Borel}{{\cal B}}
\newcommand{\Goal}{{\cal A}}
\newcommand{\Prob}[3]{{\mathbf P}^{#1}_{#2}[{#3}]}

\newcommand{\measure}[3]{{\mathbf {Pr}}^{#1}_{#2}({#3})}

\newcommand{\Paths}[1]{\operatorname{Paths}(#1)}
\newcommand{\FPaths}[1]{\operatorname{Paths}^\star(#1)}
\newcommand{\Reach}[1]{\operatorname{Reach}(#1)}
\newcommand{\FReach}[1]{\operatorname{Reach}^\star(#1)}

\newcommand{\eqdef}{\triangleq}
\newcommand{\subS}{\sqsubseteq}
\newcommand{\HsubS}{\preceq}

\newcommand{\comment}[1]{}
\newcommand{\explan}[1]{\lbrace #1 \rbrace}
\newcommand{\qdot}{\;.\;}

\newcommand{\cyl}[1]{\langle{#1}\rangle}

\newcommand{\LangPa}[1]{\operatorname{Sat}(#1)}
\newcommand{\LangSt}[1]{\operatorname{Sat}(#1)}
\newcommand{\Torrent}[1]{\operatorname{Torr}({#1})}
\newcommand{\Acyclic}[1]{\operatorname{Ac}(#1)}

\newcommand{\ce}{{\cal C}}
\newcommand{\gen}[1]{\operatorname{GenTorr}(#1)}
\newcommand{\repTorrent}[1]{\operatorname{repTorr}\left(#1\right)}
\newcommand{\tail}[1]{\operatorname{tail}(#1)}
\newcommand{\last}[1]{\operatorname{last}(#1)}
\newcommand{\SCC}{\operatorname{SCC}}
\newcommand{\SCCs}{\operatorname{SCCs}}
\newcommand{\scc}{\operatorname{K}}
\newcommand{\Inp}{\operatorname{Inp}}
\newcommand{\Out}{\operatorname{Out}}
\newcommand{\Com}{\operatorname{Com}}
\newcommand{\Distr}{\operatorname{Distr}}
\newcommand{\Sch}{\operatorname{Sch}}

\newcommand{\TorRepCount}[1]{\operatorname{TorRepCount}(#1)\xspace}
\newcommand{\pCTL}{\text{pCTL}\xspace}
\newcommand{\LTL}{\text{LTL}\xspace}

\newcommand{\MC}{\operatorname{MC}\xspace}
\newcommand{\MCs}{\operatorname{MCs}\xspace}
\newcommand{\MDP}{\operatorname{MDP}\xspace}
\newcommand{\MDPs}{\operatorname{MDPs}\xspace}
\newcommand{\SatF}[1]{\operatorname{Sat}_{\!_{\lozenge}}({#1})\xspace}

\newcommand{\mdp}{\operatorname{{\cal D}}}
\newcommand{\mc}{\operatorname{{\cal M}}}

\newcommand{\sat}[3]{\smash{{#1}\models_{_{_{{\!\!#2}}}} {#3}}}
\newcommand{\nonsat}[3]{\smash{{#1}\not\models_{_{_{\!\!{#2}}}} {#3}}}
\newcommand{\CountSet}[1]{{\cal R}({#1})}
\newcommand{\TorCountSet}[1]{{\cal R}_t({#1})}

\newtheorem{thm}{Theorem}
\numberwithin{thm}{section}
\newtheorem{lem}[thm]{Lemma}

\newtheorem{obs}[thm]{Observation}

\theoremstyle{definition}
\newtheorem{dfn}[thm]{Definition}

\begin{document}

\maketitle

\begin{abstract}
  This paper presents a novel technique for counterexample generation
  in probabilistic model checking of Markov Chains and Markov Decision
  Processes. (Finite) paths in counterexamples are grouped together
  in witnesses that are likely to provide similar debugging information
  to the user. We list five properties that witnesses should satisfy
  in order to be useful as debugging aid: similarity, accuracy, originality,
  significance, and finiteness. Our witnesses contain paths that behave
  similar outside strongly connected components.

  This papers shows how to compute these witnesses by reducing the
  problem of generating counterexamples for general properties
  over Markov Decision Processes, in several steps, to the easy
  problem of generating counterexamples for reachability properties
  over acyclic Markov Chains.
\end{abstract}

\section{Introduction}
\label{sec:intro}

%
%

Model checking is an automated technique that, given a
finite-state model of a system and a property stated in an
appropriate logical formalism, systematically checks the validity
of this property. Model checking is a general approach and is
applied in areas like hardware verification and software
engineering.

Nowadays, the interaction geometry of distributed
systems and network protocols calls for probabilistic, or more
generally, quantitative estimates of, e.g., performance and cost
measures. Randomized algorithms are increasingly utilized to
achieve high performance at the cost of obtaining correct answers
only with high probability. For all this, there is a wide
range of models and applications in computer science requiring
quantitative analysis. Probabilistic model checking allow us to
check whether or not a probabilistic property is satisfied in a given
model, e.g., ``Is every message sent successfully received with
probability greater or equal than $0.99$?''.

A major strength of model checking is the possibility of generating
diagnostic information in case the property is violated.
This diagnostic information is provided through a
\emph{counterexample} showing an execution of the model that
invalidates the property under verification.
Apart from the immediate feedback in model checking, counterexamples are also used in abstraction-refinement
techniques~\cite{clarke_2000_counterexampleguided}, and provide the foundations for schedule derivation (see, e.g.,~\cite{BLR_2005_optimalscheduling}).

%
Although counterexample generation was studied from the very
beginning in most model checking techniques, this has not been the
case for probabilistic model checking. Only recently attention was
drawn to this
subject~\cite{ahl_2005_counterexamples,al_2006_search,hk_2007_counterexamples,hk_2007_counterexamplesDTMC,al_2007_counterexamplesMDP},
fifteen years after the first studies on probabilistic model
checking.
Contrarily to other model checking techniques, counterexamples in this
setting are \emph{not} given by a single execution path.  Instead,
they are \emph{sets of executions} of the
system satisfying a certain undesired property whose probability
mass is higher than a given bound. Since counterexamples are used
as a diagnostic tool, previous works on counterexamples have
presented them as sets of finite paths of large enough
probability. We refer to these sets as \emph{representative
counterexamples}. Elements of representative counterexamples with
high probability have been considered the most informative since
they contribute mostly to the property refutation.

A challenge in counterexample generation for probabilistic model
checking is that (1) representative counterexamples are very large
(often infinite), (2) many of its elements have very low probability,
and (3) that elements can be extremely similar to each other (consequently
providing similar diagnostic information). Even worse, (4) sometimes
the finite paths with highest probability do not indicate the most
likely violation of the property under consideration.

\begin{figure}
\vspace{-0.5cm}
 \hspace{0.5cm}
 \begin{minipage}[b]{4.25cm}
   \centering
   \includegraphics[width=4.25cm]{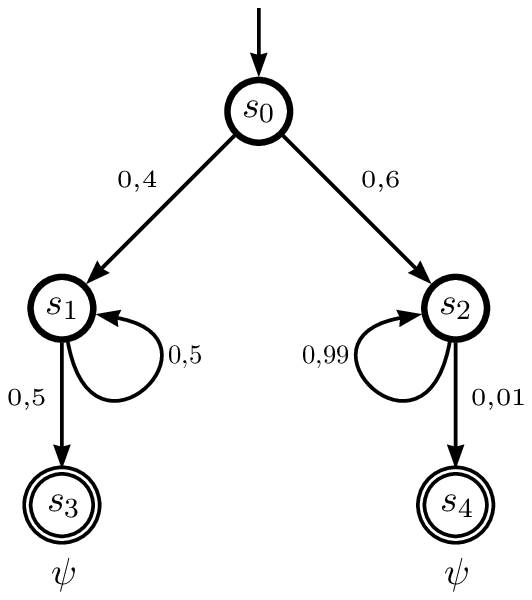}\caption{Markov chain}\label{fig:MC related to MDP}
 \end{minipage}
 \hspace{1.5cm}
 \begin{minipage}[b]{4 cm}
   \begin{tabular}{|c||c|c||c|c|}
   \hline \multicolumn{1}{|c||}{}&\multicolumn{2}{|c||}{\textbf{Single paths}}&\multicolumn{2}{|c|}{\textbf{Witnesses}}\\
   \hline Rank&F. Path&Prob&Witness&Mass\\
   \hline
   1&$s_0(s_1)^1s_3$&0.2&[$s_0 s_2 s_4$]&0.6\\
   2&$s_0(s_1)^2s_3$&0.1&[$s_0 s_1 s_3$]&0.4\\
   3&$s_0(s_1)^3s_3$&0.05&&\\
   4&$s_0(s_1)^4s_3$&0.025&&\\
   5&$s_0(s_1)^5s_3$&0.0125&&\\
   6&$s_0(s_1)^6s_3$&0.00625&&\\
   7&$s_0(s_2)^1s_4$&0.006&&\\
   8&$s_0(s_2)^2s_4$&0.0059&&\\
   9&$s_0(s_2)^3s_4$&0.0058&&\\
   \vdots&\vdots&\vdots&&\\
   \hline
   \end{tabular}
   \caption{\ \ Comparison Table}\label{tbl:comparison}
 \end{minipage}
\vspace{-0.3cm}
\end{figure}

For example, look at the Markov chain $\mc$ in
Figure~\ref{fig:MC related to MDP}.
The property $\sat{\mc}{\leq 0.5}{\F \psi}$ stating that execution
reaches a state satisfying $\psi$ (i.e., reaches $s_3$ or $s_4$)
with probability lower or equal than $0.5$ is violated (since the
probability of reaching $\psi$ is 1). The left hand side of table
in Figure~\ref{tbl:comparison} lists finite paths reaching $\psi$
ranked according to their probability. Note that finite paths with
highest probability take the left branch in the system, whereas
the right branch in itself has higher probability, illustrating
Problem 4. To adjust the model so that it does satisfy the
property (bug fixing), it is not sufficient to modify the left
hand side of the system alone; no matter how one changes the left
hand side, the probability of reaching $\psi$ remains at least
$0.6$. Furthermore, the first six finite paths provide similar
diagnostic information: they just make extra loops in $s_1$. This
is an example of Problem 3. Also, the probability of every single
finite path is far below the bound $0.5$, making it unclear if a
particular path is important; see Problem 2 above. Finally, the
(unique) counterexample for the property $\sat{\mc}{< 1}{\F \psi}$
consists of infinitely many finite paths (namely all finite paths
of $\mc$); see Problem 1.
To overcome these problems, we partition a representative
counterexample
into sets of finite paths that follow a 
similar pattern.
We call these sets \emph{witnesses}.
%
To ensure that witnesses provide valuable diagnostic information,
we desire that the set of witnesses that form a
counterexample satisfies several properties:
two different witnesses should provide different diagnostic
information (solving Problem 3) and elements of a single witness
should provide similar diagnostic information, as a consequence
witnesses have a high probability mass (solving Problems 2 and 4),
and the number of witnesses of a representative counterexample
should be finite (solving Problem 1).

In our setting, witnesses consist of paths that behave the same
outside strongly connected components. In the example of
Figure~\ref{fig:MC related to MDP}, there are two witnesses: the
set of all finite paths going right, represented by [$s_0 s_2
s_4$] whose probability (mass) is $0.6$, and the set of all finite
paths going left, represented by [$s_0 s_1 s_3$] with probability
(mass) $0.4$.





In this paper, we show how to obtain such sets of witnesses for bounded
probabilistic LTL properties on Markov decision processes ($\MDP$).
In fact, we first show how to reduce this problem to finding witnesses
for upper bounded probabilistic reachability properties on discrete
time Markov chains ($\MCs$).
The major technical matters lie on this last problem to which most of
the paper is devoted.


In a nutshell, the process to find witnesses for the violation of
$\sat{\mc}{\leq p}{\F \psi}$, with $\mc$ being a $\MC$, is as
follows.
We first eliminate from the original $\MC$ all the
``uninteresting'' parts.  This proceeds as the first steps of the
model checking process: make absorbing all state satisfying
$\psi$, and all states that cannot reach $\psi$, obtaining a new
$\MC$ $\mc_\psi$.  Next reduce this last $\MC$ to an acyclic $\MC$
$\Acyclic{\mc_\psi}$ in which all strongly connected components
have been conveniently abstracted with a single probabilistic
transition.
The original and the acyclic $\MC$s are related by a mapping that,
to each finite path in $\Acyclic{\mc_\psi}$ (that we call rail),
assigns a set of finite paths behaving similarly in $\mc$ (that we
call torrent).
This map preserves the probability of reaching $\psi$ and hence
relates counterexamples in $\Acyclic{\mc_\psi}$ to counterexamples
in $\mc$.
Finally, counterexamples in $\Acyclic{\mc_\psi}$ are computed by
reducing the problem to a $k$ shortest path problem, as in
\cite{hk_2007_counterexamples}. Because $\Acyclic{\mc_\psi}$ is
acyclic, the complexity is lower than the corresponding problem in
\cite{hk_2007_counterexamples}.

It is worth to mention that our technique can also be applied to
simple $\pCTL$ formulas without nested path quantifiers.


\paragraph*{Organization of the paper.}
Section~\ref{sec:preliminaries} presents the necessary background
on Markov chains ($\MC$), Markov Decision Processes ($\MDP$), and
Linear Temporal Logic ($\LTL$).
Section~\ref{sec:counterexamples} presents the definition of
counterexamples and discuss the reduction from general $\LTL$
formulas to upper bounded probabilistic reachability properties,
and the extraction of the maximizing $\MC$ in a $\MDP$.
Section~\ref{sec:RepCount-parti-witnesses} discusses desire
properties of counterexamples.  In
Sections~\ref{sec:Torrents-Rails}
and~\ref{sec:significantdiagnosticcounterexamples}, we introduce
the fundamentals on rails and torrents, the reduction of the
original $\MC$ to the acyclic one, and our notion of significant
diagnostic counterexamples.
Section~\ref{sec:coumputing-counterexamples} then present the
techniques to actually compute counterexamples. In
Section~\ref{sec:conclusions} we discuss related work and give final
conclusions.

\section{Preliminaries}
\label{sec:preliminaries}

\subsection{Markov Decision Processes and Markov chains}

Markov Decision Processes ($\MDPs$) constitute a formalism that
combines nondeterministic and probabilistic choices.  They are the
dominant model in corporate finance, supply chain optimization,
system verification and optimization.  There are many slightly
different variants of this formalism such as action-labeled
$\MDPs$ \cite{bel_1957_mdp,FilarVrieze97}, probabilistic automata
\cite{sl_1995_probabilistic,sv_2004_automata}; we work with the
state-labeled $\MDPs$ from~\cite{ba_1995_probabilistic}.

\begin{dfn} Let $S$ be a set. A \emph{discrete probability
distribution} on $S$ is a function $p \colon S \to [0,1]$ with
countable or finite carrier and such that $\sum_{s\in S}p(s)=1$.
We denote the set of all discrete probability distributions on $S$
by $\Distr(S)$.
Additionally, we define the \emph{Dirac distribution} on an
element $s \in S$ as $1_s$, i.e., $1_s(s)=1$ and $1_s(t)=0$ for
all $t\in S\setminus\lbrace s \rbrace$.
\end{dfn}

\begin{dfn}\label{dfn:mdp}
  A \emph{Markov Decision Process} ($\MDP$) is a four-tuple
  $\mdp=(S,s_0,L,\tau)$, where
\begin{itemize}
  \item $S$ is the finite state space of the system;
  \item $s_0 \in S$ is the initial state;
  \item $L$ is a labeling function that associates to each state $s\in
    S$ a set $L(s)$ of propositional variables that are \emph{valid} in $s$;
  \item $\tau \colon S \to \wp(\Distr(S))$ is a function that
  associates to each $s \in S$ a non-empty and finite subset of
  $\Distr(S)$ of probability distributions.
\end{itemize}
\end{dfn}

\begin{dfn} Let $\mdp=(S,s_0,\tau,L)$ be a $\MDP$. We define a \emph{successor} relation $\delta\subseteq
S\times S$ by $\delta \eqdef \{(s,t)|\exists\, \pi\in\tau(s) \qdot
\pi(t)>0\}$ and for each state $s\in S$ we define the sets
\begin{align*}
 \Paths{\mdp, s}  \eqdef \{s_0s_1s_2\ldots\in S^\omega | s_0=s\land\forall n\in \mathbb{N} \qdot
\delta(s_n,s_{n+1})\}\mbox{ and}\\
 \FPaths{\mdp, s} \eqdef \{s_0s_1\ldots s_n\in S^\star | s_0=s\land\forall\, 0\leq i < n \qdot \delta(s_n,s_{n+1})\}
\end{align*}
  of paths and finite paths respectively beginning at $s$. We usually omit $\mdp$ from the notation; we also abbreviate $\Paths{\mdp, s_0}$ as $\Paths{\mdp}$ and $\FPaths{\mdp, s_0}$ as $\FPaths{\mdp}$. For $\omega
  \in\Paths{s}$, we write the $(n\!+\!1)$-st state of $\omega$ as $\omega_n$.
  As usual, we let $\Borel_s\subseteq \wp(\Paths{s})$ be the Borel
  $\sigma$-algebra on the cones $\cyl{s_0 \dots s_n} \eqdef
  \{\omega \in \Paths{s} | \omega_0=s_0 \land \ldots \land
  \omega_n=s_n\}$. Additionally, for a set of finite paths $\Lambda\subseteq\FPaths{s}$, we define $\cyl{\Lambda}\eqdef
\bigcup_{\sigma\in\Lambda}\cyl{\sigma}$.
\end{dfn}
\setlength{\abovecaptionskip}{-6pt plus 1pt minus 1pt}
\vspace{-1cm}
\begin{figure}
    \begin{center}
      \includegraphics[width=8cm]{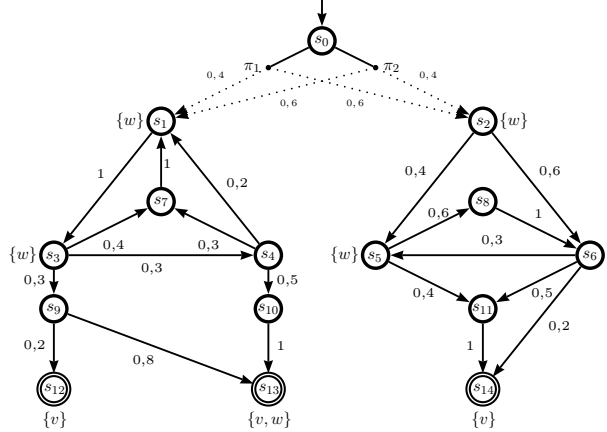}
    \end{center}
    \caption{Markov Decision Process}
    \label{fig:notMaxNodes}
\end{figure}
\vspace{-0,25cm} \label{exa:MDP} Figure~\ref{fig:notMaxNodes}
shows a $\MDP$. Absorbing states (i.e., states $s$ with
$\tau(s)=\{1_s\}$) are represented by double lines.  This $\MDP$
features a single nondeterministic decision, to be made in state
$s_0$, namely $\pi_1$ and $\pi_2$.

\begin{dfn} Let $\mdp=(S,s_0,\tau,L)$ be a $\MDP$ and $\Goal\subseteq S$. We define the sets of paths and finite paths
reaching  $\Goal$ as
\begin{align*}
\Reach{\mdp,s,\Goal}\triangleq\lbrace \omega\in \Paths{\mdp,s}\mid
\exists_{i\geq 0}. \omega_i\in\Goal\rbrace\mbox{ and}\\
\FReach{\mdp,s,\Goal}\triangleq\lbrace\sigma\in\FPaths{\mdp,s}\mid
\last{\sigma}\in \Goal \land \forall_{i\leq
|\sigma|-1}.\sigma_i\not \in \Goal\rbrace
\end{align*}
respectively. Note that $\FReach{\mdp,s,\Goal}$ consists
of those finite paths $\sigma$ reaching $\Goal$ exactly once, at
the end of the execution. It is easy to check that these sets are
\emph{prefix free}, i.e. contain finite paths such that none of
them is a prefix of another one.
\end{dfn}

\subsection{Schedulers}

Schedulers (also called strategies, adversaries, or policies)
resolve the nondeterministic choices in a
$\MDP$~\cite{pz_1993_verification,var_1985_probabilistic,ba_1995_probabilistic}.

\begin{dfn}\label{dfn:scheduler}
  Let $\mdp=(S,s_0,\tau,L)$ be a $\MDP$. A \emph{scheduler} $\eta$ on $\mdp$ is a function from $\FPaths{\mdp}$
  to $\Distr(\wp(\Distr(S)))$ such that for all $\sigma\in \FPaths{\mdp}$
  we have $\eta(\sigma)\in \Distr(\tau(\last{\sigma}))$. We denote
  the set of all schedulers on $\mdp$ by $\Sch(\mdp)$.
\end{dfn}


Note that our schedulers are randomized, i.e., in a
finite path $\sigma$ a scheduler chooses an element of
$\tau(\last{\sigma})$ probabilistically.  Under a scheduler
$\eta$, the probability that the next state reached after the path
$\sigma$ is $t$, equals $\sum_{\pi\in
  \tau(\last{\sigma})}\eta(\sigma)(\pi) \cdot \pi(t)$. In this way, a
scheduler induces a probability measure on $\Borel_s$ as usual.

\begin{dfn}
  Let $\mdp$ be a $\MDP$, $s\in S$, and $\eta$ an $s$-scheduler on
  $\mdp$. We define the probability measure $\mu_{s,\eta}$ as the
  unique measure on $\Borel_s$ such that for all $s_0 s_1\ldots s_n\in
  \FPaths{s}$
\begin{align*}\measure{}{s,\eta}{\cyl{s_0s_1\ldots s_n}}
    = \prod_{i=0}^{n-1} \sum_{\pi\in \tau(s_i)}
    \eta(s_0 s_1\ldots s_i)(\pi) \cdot \pi(s_{i+1}).
\end{align*}
\end{dfn}
We now recall the notions of deterministic and memoryless
schedulers.

\begin{dfn}
  Let $\mdp$ be a $\MDP$, $s\in S$, and $\eta$ an scheduler of $\mdp$. We
  say that $\eta$ is \emph{deterministic} if $\eta(\sigma)(\pi_i)$ is
  either $0$ or $1$ for all $\pi_i\in\tau(\last{\sigma})$ and all $\sigma
  \in\FPaths{\mdp}$.
  We say that a scheduler is \emph{memoryless} if for all finite paths $\sigma_1,\sigma_2$ of $\mdp$ with
  $\last{\sigma_1}=\last{\sigma_2}$ we have
  $\eta(\sigma_1)=\eta(\sigma_2)$
\end{dfn}

\begin{dfn}
  Let $\mdp$ be a $\MDP$, $s\in S$, and $\Delta \in \Borel_s$. Then the
  \emph{maximal and minimal probabilities of $\Delta$},
  $\measure{+}{s}{\Delta}, \measure{-}{s}{\Delta}$, are defined by
  $$
    \measure{+}{s}{\Delta}\eqdef \sup_{\eta\in \Sch_s(\mdp)}
\measure{}{s,\eta}{\Delta}
    \hspace{0.5cm} \text{and} \hspace{0.5cm}
    \measure{-}{s}{\Delta}\eqdef \inf_{\eta\in \Sch_s(\mdp)} \measure{}{s,\eta}{\Delta}.
  $$
  A scheduler that attains $\measure{+}{s}{\Delta}$ or $\measure{-}{s}{\Delta}$
  is called a \emph{maximizing} or \emph{minimizing} scheduler
  respectively.
\end{dfn}

\noindent
A \emph{Markov chain} ($\MC$) is a $\MDP$ associating exactly one
probability distribution to each state. In this way
nondeterministic choices are not longer allowed.

\begin{dfn}[Markov chain] Let $\mdp=(S,s_0,\tau,L)$ be a $\MDP$.
If $|\tau(s)|=1$ for all $s\in S$, then we say that $\mdp$ is a
\emph{Markov chain} ($\MC$).
\end{dfn}

\subsection{Linear Temporal Logic}

Linear temporal logic ($\LTL$) \cite{MP_1991_LTL} is a modal
temporal logic with modalities referring to time. In $\LTL$ is
possible to encode formulas about the future of paths: a condition
will eventually be true, a condition will be true until another
fact becomes true, etc.

\begin{dfn} $\LTL$ is built up from the set of propositional variables $\Var$, the logical connectives $\lnot$,
$\land$, and a temporal modal operator by the following grammar:
$$\phi ::= \Var \mid \lnot \phi \mid \phi\land \phi \mid \Until{\phi}{\phi}.$$
\noindent Using these operators we define $\lor,\rightarrow,\F{},$
and $\G{}$ in the standard way.
\end{dfn}

\begin{dfn} Let $\mdp=(S,s_0, \tau,L)$ be a $\MDP$. We define satisfiability for paths $\omega$ in $\mdp$ and
$\LTL$ formulas $\phi,\psi$ inductively by
$$
\begin{array}{lcl}
\sat{\omega}{\mdp}{v} &\Leftrightarrow & v\in L(\omega_0)\\
\sat{\omega}{\mdp}{\lnot \phi} &\Leftrightarrow & \mbox{not}(\sat{\omega}{\mdp}{\phi})\\
\sat{\omega}{\mdp}{\phi\land\psi} &\Leftrightarrow & \sat{\omega}{\mdp}{\phi} \mbox{ and }\sat{\omega}{\mdp}{\psi}\\
\sat{\omega}{\mdp}{\Until{\phi}{\psi}} & \Leftrightarrow & \exists_{i\geq 0}. \sat{\omega_{\downarrow i}}{\mdp}{\psi}\mbox{ and }\forall_{0\leq j < i}.\sat{\omega_{\downarrow j}}{\mdp}{\phi}
\end{array}
$$
where $\omega_{\downarrow i}$ is the $i$-th suffix of
$\omega$. When confusion is unlikely, we omit the subscript $\mdp$
on the satisfiability relation.
\end{dfn}

\begin{dfn} Let $\mdp$ be a $\MDP$. We define the language $\text{Sat}_{_{\mdp}}(\phi)$ associated to an $\LTL$ formula $\phi$
as the set of paths satisfying $\phi$, i.e.
$\text{Sat}_{_{\mdp}}(\phi)\triangleq \lbrace
\omega\in\Paths{\mdp}\mid \omega\models \phi\rbrace.$ Here we also
generally omit the subscript $\mdp$.
\end{dfn}

We now define satisfiability of an $\LTL$ formula $\phi$ on a
$\MDP$ $\mdp$. We say that $\mdp$ satisfies $\phi$ with
probability at most $p$ ($\sat{\mdp}{\leq p}{\phi}$) if the
probability of getting an execution satisfying $\phi$ is at most
$p$.

\begin{dfn} Let $\mdp$ be a $\MDP$, $\phi$ an $\LTL$ formula and $p\in[0,1]$. We define $\sat{}{\leq p}{}$ and $\sat{}{\geq p}{}$by
\begin{align*}
\sat{\mdp}{\leq p}{\phi} \Leftrightarrow \measure{+}{_{\mdp}}{\LangPa{\phi}}\leq p,\\
\sat{\mdp}{\geq p}{\phi} \Leftrightarrow
\measure{-}{_{\mdp}}{\LangPa{\phi}}\geq p.
\end{align*}
\noindent We define $\sat{\mdp}{< p}{\phi}$ and $\sat{\mdp}{>
p}{\phi}$ in a similar way.
\end{dfn}

In case the $\MDP$ is fully probabilistic, i.e., a $\MC$, the
satisfiability problem is reduced to $\sat{\mdp}{\bowtie p}{\phi}
\Leftrightarrow \measure{}{_{\mdp}}{\LangPa{\phi}}\bowtie p$,
where $\bowtie\in \lbrace <,\leq,>,\geq\rbrace$.

\section{Counterexamples}\label{sec:counterexamples}

In this section, we define what counterexamples are and how
the problem of finding counterexamples for a general $\LTL$ property
over Markov Decision Processes reduces to finding counterexamples
to reachability problems over Markov chains.

\begin{dfn}[Counterexamples]\label{dfn:counterexamples}%
  Let $\mdp$ be a $\MDP$ and $\phi$ an $\LTL$ formula. A
  \emph{counterexample} to $\sat{\mdp}{\leq p}{\phi}$ is a
  measurable set $\ce\subseteq \LangPa{\phi}$ such that
  $\measure{+}{\mdp}{\ce}> p$.
  Counterexamples to $\sat{\mdp}{< p}{\phi}$ are defined similarly.
\end{dfn}

Counterexamples to $\sat{\mdp}{> p}{\phi}$ and $\sat{\mdp}{\geq
  p}{\phi}$ cannot be defined straightforwardly as it is always
possible to find a set $\ce\subseteq \LangPa{\phi}$ such that
$\measure{-}{\mdp}{\ce}\leq p$ or $\measure{-}{\mdp}{\ce}< p$,
note that the empty set trivially satisfies it.
Therefore, the best way to find counterexamples to lower bounded
probabilities is to find counterexamples to the dual properties
$\sat{\mdp}{< 1-p}{\!\!\!\neg\phi}$ and $\sat{\mdp}{\leq
  1-p}{\!\!\!\neg\phi}$. 
That is, while for upper bounded probabilities, a counterexample
is a set of paths satisfying the property beyond the bound, for
lower bounded probabilities the counterexample is a set of paths
that \emph{does not} satisfy the property with sufficient
probability.

\begin{wrapfigure}{r}{5.2cm}
\vspace{-1.25cm}
\includegraphics[width=5cm]{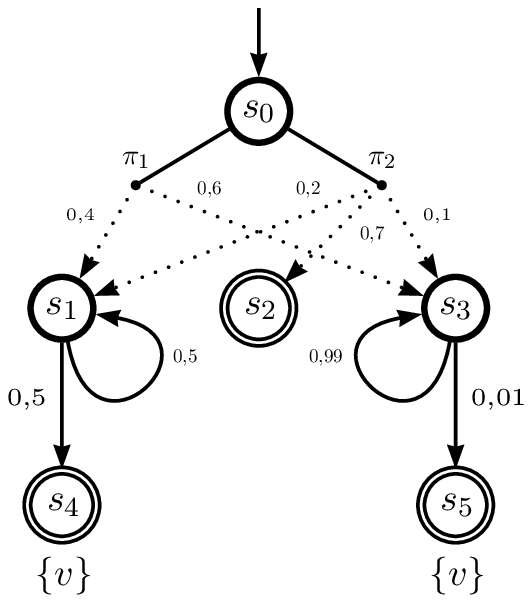}
\caption{} \label{fig:ExaModelCheckerCuant} \vspace{-6ex}
\end{wrapfigure}
%
\medskip\par\noindent%
\textit{Example~\refstepcounter{example}\label{exa:counterexamples}\ref{exa:counterexamples}.}
  Consider the $\MDP$ $\mdp$ of Figure~\ref{fig:ExaModelCheckerCuant}
  and the $\LTL$ formula $\F v$. it is easy to check that
  $\nonsat{\mdp}{< 1}{\F v}$. The set $\ce=\LangPa{\F
    v}=\{\gamma\!\in\! \Paths{s_0}|\exists_{i\geq 0}.\gamma\!=\!s_0
  (s_1)^i (s_4)^\omega\}\cup\{\gamma\!\in\!  \Paths{s_0}|\exists_{i\geq
    0}.\gamma\!=\!s_0 (s_3)^i (s_5)^\omega\}$ is a counterexample. Note
  that $\measure{+}{s_0,\eta}{\ce}\!=\!1$ where $\eta$ is any
  deterministic scheduler of $\mc$ satisfying $\eta(s_0)=\pi_1$.
\medskip

$\LTL$ formulas are actually checked by reducing the model
checking problem to a reachability
problem~\cite{dealfaro_1997_hybrid}. For checking upper bounded
probabilities, the $\LTL$ formula is translated into an equivalent
deterministic Rabin automaton and composed with the $\MDP$ under
verification. On the obtained $\MDP$, the set of states forming
accepting end components (maximal components that traps accepting
conditions with probability 1) are identified.
The maximum probability of the $\LTL$ property on the original
$\MDP$ is the same as the maximum probability of reaching a state
of an accepting end component in the final $\MDP$.
Hence, from now on we will focus on counterexamples to properties
of the form $\sat{\mdp}{\leq p}{\F\psi}$ or $\sat{\mdp}{<
p}{\F\psi}$, where $\psi$ is a propositional formula, i.e., a
formula without temporal operators.

In the following, it will be useful to identify the set of states
in which a propositional property is valid.

\begin{dfn}%
  Let $\mdp$ be a $\MDP$. We define the state language
  $\text{Sat}_{\mdp}(\psi)$ associated to a propositional formula $\psi$
  as the set of states satisfying $\psi$,
  i.e., $\text{Sat}_{\mdp}(\psi)\triangleq \lbrace s\in S\mid s\models\psi\rbrace$,
  where $\models$ has the obvious satisfaction meaning for states. As
  usual, we generally omit the subscript $\mdp$.
\end{dfn}

To find a counterexample to a property in a $\MDP$ with respect to
a upper bound, it suffices to find a counterexample for the
maximizing scheduler. A scheduler defines a Markov chain, and
hence finding a counterexample on a $\MDP$ amounts to finding a
counterexample in the Markov chain induced by the maximizing
scheduler.
The maximizing scheduler turns out to be deterministic and
memoryless~\cite{ba_1995_probabilistic}; consequently the induced
Markov chain can be easily extracted from the $\MDP$ as follows.

\begin{dfn}%
  Let $\mdp=(S,s_0,\tau,L)$ be a $\MDP$ and $\eta$ a deterministic
  memoryless scheduler. Then we define the $\MC$ $\eta$-associated to
  $\mdp$ as $\mdp_\eta=(S,s_0,\M_\eta,L)$ where
  $\M_{\eta}(s,t)=(\eta(s))(t)$ for all $s,t\in S$.
\end{dfn}

Now we state that finding counterexamples for upper bounded
probabilistic reachability $\LTL$ properties on $\MDPs$ can be
reduced to finding counterexamples for upper bounded probabilistic
reachability $\LTL$ properties on $\MCs$.

\begin{thm}%
  Let $\mdp$ be a $\MDP$, $\psi$ a propositional formula and
  $p\in[0,1]$.  Then, there is a maximizing (deterministic memoryless)
  scheduler $\eta$ such that
  $\sat{\mdp}{\leq p}{\F\psi}\Leftrightarrow \sat{\mdp_\eta}{\leq p}{\F\psi}$.
  Moreover,
  $\ce$ is a counterexample to $\sat{\mdp_\eta}{\leq p}{\F\psi}$ if
  and only if $\ce$ is also a counterexample to $\sat{\mdp}{\leq p}{\F\psi}$.
\end{thm}

\section{Representative Counterexamples, Partitions and
Witnesses}\label{sec:RepCount-parti-witnesses}

The notion of counterexample from Definition
\ref{dfn:counterexamples} is very broad: just an arbitrary
(measurable) set of paths with high enough probability. To be
useful as a debugging tool (and in fact to be able to present the
counterexample to a user), we need counterexamples with specific
properties. We will partition counterexamples (or rather, representative counterexamples) in witnesses and list five properties that witnesses should satisfy.

The first point to stress is that for reachability properties it
is sufficient to consider counterexamples that consist of finite
paths.

\begin{dfn}[Representative counterexamples] Let $\mdp$ be a $\MDP$, $\psi$
a propositional formula and $p\in[0,1]$. A \emph{representative
counterexample} to $\sat{\mdp}{\leq p}{\F \psi}$ is a set
$\ce\subseteq \FReach{\mdp,\LangSt{\psi}}$ such that
$\measure{+}{\mdp}{\cyl{\ce}}>p$. We denote the set of all
representative counterexamples to $\sat{\mc}{\leq p}{\F \psi}$ by
$\CountSet{\mc, p,\psi}$.
\end{dfn}

\begin{thm} Let $\mdp$ be a $\MDP$, $\psi$ a propositional formula and $p\in[0,1]$. If $\ce$ is a representative counterexample to $\sat{\mdp}{\leq p}{\F
\psi}$, then $\cyl{\ce}$ is a counterexample to $\sat{\mdp}{\leq
p}{\F\psi}$. Furthermore, there exists a counterexample to
$\sat{\mdp}{\leq p}{\F\psi}$ if and only if there exists a
representative counterexample to $\sat{\mdp}{\leq p}{\psi}$.
\end{thm}

Following \cite{hk_2007_counterexamples}, we present the notions
of \emph{minimum counterexample}, \emph{strongest evidence} and
\emph{most indicative counterexamples}.

\begin{dfn}[Minimum counterexample] Let $\mc$ be a $\MC$, $\psi$ a propositional formula and $p\in[0,1]$. We say that $\ce\in
\CountSet{\mc,p,\psi}$ is a \emph{minimum counterexample} if
$|\ce|\leq|\ce^\prime|$, for all
$\ce^\prime\in\CountSet{\mc,p,\psi}$.
\end{dfn}

\begin{dfn}[Strongest evidence] Let $\mc$ be a $\MC$, $\psi$ a propositional formula and $p\in[0,1]$. A \emph{strongest evidence} to $\smash{\nonsat{\mc}{\leq p}{\F \psi}}$ is a finite path $\sigma\in \FReach{\mc,\LangSt{\psi}}$ such that $\measure{}{\mc}{\cyl{\sigma}}\geq\measure{}{\mc}{\cyl{\rho}}$, for all $\rho \in\FReach{\mc,\LangSt{\psi}}$.
\end{dfn}

\begin{dfn}[Most indicative counterexample] Let $\mc$ be a $\MC$, $\psi$ a propositional formula and
$p\in[0,1]$. We call $\ce\in \CountSet{\mc,p,\psi}$ a \emph{most
indicative counterexample} if it is minimum and
$\measure{}{}{\cyl{\ce}}\geq \measure{}{}{\cyl{\ce^\prime}}$, for
all minimum counterexamples $\ce^\prime \in
\CountSet{\mc,p,\psi}$.
\end{dfn}

Unfortunately, very often most indicative counterexamples are very
large (even infinite), many of its elements have insignificant
measure and elements can be extremely similar to each other
(consequently providing the same diagnostic information). Even
worse, sometimes the finite paths with highest probability do not
exhibit the way in which the system accumulates higher probability
to reach the undesired property (and consequently where an error
occurs with higher probability). For these reasons, we are of the
opinion that representative counterexamples are still too general
in order to be useful as feedback information. We approach this
problem by splitting out the representative counterexample into
sets of finite paths following a ``similarity'' criteria
(introduced in Section~\ref{sec:Torrents-Rails}). These sets are
called \emph{witnesses of the counterexample}.


Recall that a set $Y$ of nonempty sets is a partition of $X$ if
the elements of $Y$ cover $X$ and the elements of $Y$ are pairwise
disjoint. We define counterexample partitions in the following
way.

\begin{dfn}[Counterexample partitions and witnesses] Let $\mdp$ be
a $\MDP$, $\psi$ a propositional formula, $p\in[0,1]$, and $\ce$
a representative counterexample to $\sat{\mdp}{\leq p}{\F
\psi}$. A \emph{counterexample partition} $W_\ce$ is a partition
of $\ce$. We call the elements of $W_\ce$ \emph{witnesses}.
\end{dfn}

Since not every partition generates useful witnesses (from the
debugging perspective), we now state properties that witnesses
must satisfy in order to be valuable as diagnostic information. In
Section \ref{sec:coumputing-counterexamples} we show how to
partition the detailed counterexample in order to obtain useful
witnesses.

\begin{itemize}
\item[\textbf{Similarity:}]{Elements of a witness should provide similar debugging information.}
\item[\textbf{Accuracy:}]{Witnesses with higher probability should show evolution of the system with higher probability of containing errors.}
\item[\textbf{Originality:}]{Different witnesses should provide different debugging information.}
\item[\textbf{Significance:}]{The probability of a witnesses should be close to the probability bound $p$.}
\item[\textbf{Finiteness:}]{The number of witnesses of a counterexamples partition should be finite.}
\end{itemize}

\section{Rails and Torrents}\label{sec:Torrents-Rails}

As argued before we consider that representative counterexamples
are excessively general to be useful as feedback information.
Therefore, we group finite paths of a representative
counterexample in witnesses if they are ``similar enough''. We
will consider finite paths that behave the same outside $\SCCs$ of
the system as providing similar feedback information.

In order to formalize this idea, we first reduce the original
Markov chain to an acyclic one that preserves reachability
probabilities. We do so by removing all $\SCCs$ $\scc$ of $\mc$
keeping just \emph{input states} of $\scc$. In this way, we get a
new acyclic $\MC$ denoted by $\Acyclic{\mc}$. The probability
matrix of the Markov chain relates input states of each $\SCC$
with its \emph{output states} with the reachability probability
between these states in $\mc$. Secondly, we establish a map
between finite paths $\sigma$ in $\Acyclic{\mc}$ (\emph{rails})
and sets of finite paths $W_\sigma$ in $\mc$ (\emph{torrents}).
Each torrent contains finite paths that are similar, i.e., behave
the same outside $\SCCs$. Additionally we show that the
probability of $\sigma$ is equal to the probability of $W_\sigma$.

\subsection*{Reduction to Acyclic Markov Chains}

Consider a $\MC$ $\mc=(S,s_0, \M,L)$. Recall that a subset
$\scc\subseteq S$ is called \emph{strongly connected} if for every
$s,t\in \scc$ there is a finite path from $s$ to $t$. Additionally
$\scc$ is called a \emph{strongly connected component} ($\SCC$) if
it is a maximally (with respect to $\subseteq$) strongly connected
subset of $S$.

Note that every state is a member of
exactly one $\SCC$ of $\mc$ (even those states that are not
involved in cycles, since the trivial finite path $s$ connects $s$
to itself). From now on we let $\SCC^\star$ be the set of non
trivial strongly connected components of a $\MC$, i.e., those
composed of more than one state.

A Markov chain is called \emph{acyclic} if it does not have non
trivial $\SCCs$. Note that an acyclic Markov chain still has
absorbing states.

\begin{dfn}
Let $\mc=(S,s_0,\M,L)$ be a $\MC$. Then, for each $\SCC^\star$
$\scc$ of $\mc$, we define the sets $\Inp_{\scc}\subseteq S$ of
all states in $\scc$ that have an incoming transition from a state
outside of $\scc$ and $\Out_{\scc}\subseteq S$ of all states
outside of $\scc$ that have an incoming transition from a state of
$\scc$ in the following way

\begin{figure}[hbt]
\vspace{-0.8cm} \hspace{0,5 cm}
\begin{minipage}[t]{6cm}
  \vspace{-2.5cm}
  \centering
  \begin{align*}
  \Inp_{\scc}\triangleq\{u\in \scc\mid\exists\,s\in S\setminus\scc.
  \M(s,u)>0\},\\
  \Out_{\scc}\triangleq\{s\in S\setminus\scc\mid\exists\,u\in \scc.
  \M(u,s)>0\}.\\
  \end{align*}
\end{minipage}
\hspace{1,4 cm} \vspace{-0.5cm}
\begin{minipage}[t]{5cm}
  \centering
  \includegraphics[width=3cm]{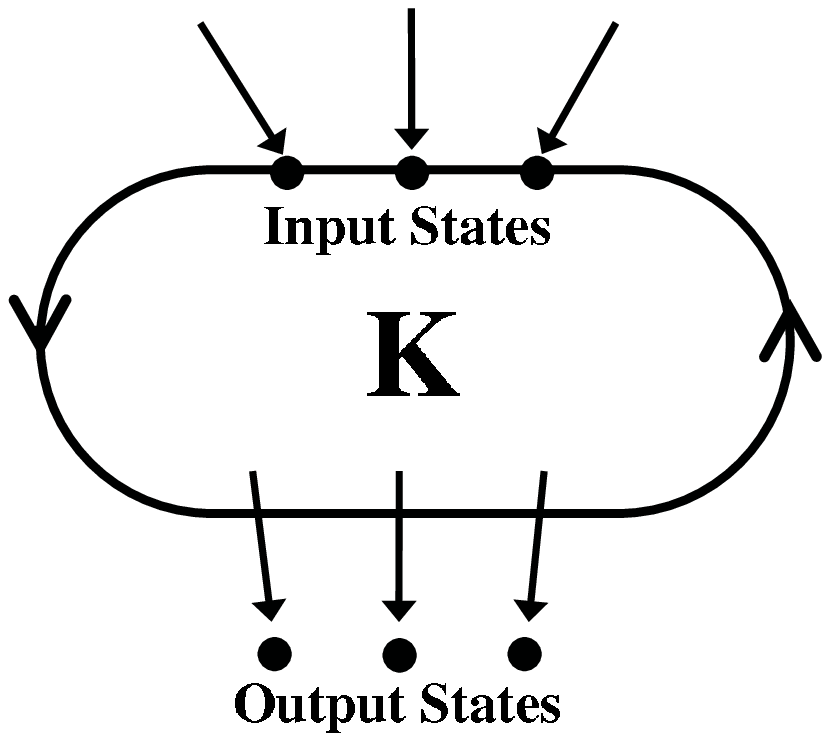}
\end{minipage}
\vspace{-1cm}
\end{figure}

\noindent We also define for each $\SCC^\star$ $\scc$ a $\MC$
related to $\scc$ as $\mc_{\scc}\triangleq(\scc\cup
\Out_{\scc},s_{\scc},\M_{\scc},L_{\scc})$ where $s_{\scc}$ is any
state in $\Inp_{\scc}$, $L_{\scc}(s)\triangleq L(s)$, and
$\M_{\scc}(s,t)$ is equal to $\M(s,t)$ if $s\in \scc$ and equal to
$1_s$ otherwise. Additionally, for every state $s$ involved in non
trivial $\SCCs$ we define $\SCC^+_s$ as $\mc_{\scc}$, where $\scc$
is the $\SCC^\star$ of $\mc$ such that $s\in \scc$.
\end{dfn}

Now we are able to define an acyclic $\MC$ $\Acyclic{\mc}$ related
to $\mc$.

\begin{dfn}\label{dfn:acyclicMC}
Let $\mc=(S,s_0,\M,L)$ be a  $\MC$. We define
$\Acyclic{\mc}{}\triangleq(S^\prime,s_0,\M^\prime{},L^\prime)$
where
\vspace{-0.5cm}
\begin{itemize}
\item{$S^\prime\triangleq\stackrel{S_{\text{com}}}{\overbrace{S\setminus\bigcup_{\scc\in \SCC^\star} \scc}}\bigcup\stackrel{S_{\text{inp}}}{\overbrace{\bigcup_{\scc\in \SCC^\star} \Inp_{\scc}}}$}
\item $L^\prime\triangleq L_{|_{S^\prime}}$,
\item $ \M^\prime(s,t) \triangleq \left\lbrace
                           \begin{array}{ll}
                             \M(s,t) & \mbox{if } s\in S_{com},\\
                             \measure{}{_{\mc,s}}{\Reach{\SCC^+_s,s,\lbrace t\rbrace}} & \mbox{if } s\in S_{inp}\land t\in \Out_{\SCC^+_s},\\
                             1_s & \mbox{if } s\in S_{inp}\land \Out_{\SCC^+_s}=\emptyset,\\
                             0 & \mbox{otherwise.}
                             \end{array}
              \right.$
\end{itemize}
\end{dfn}

Note that $\Acyclic{\mc}$ is indeed acyclic.

\begin{figure}[h]
 \vspace{-0.5cm}
 \centering
  \subfigure[Original $\MC$]{\label{fig:red-a}\includegraphics[width=6cm]{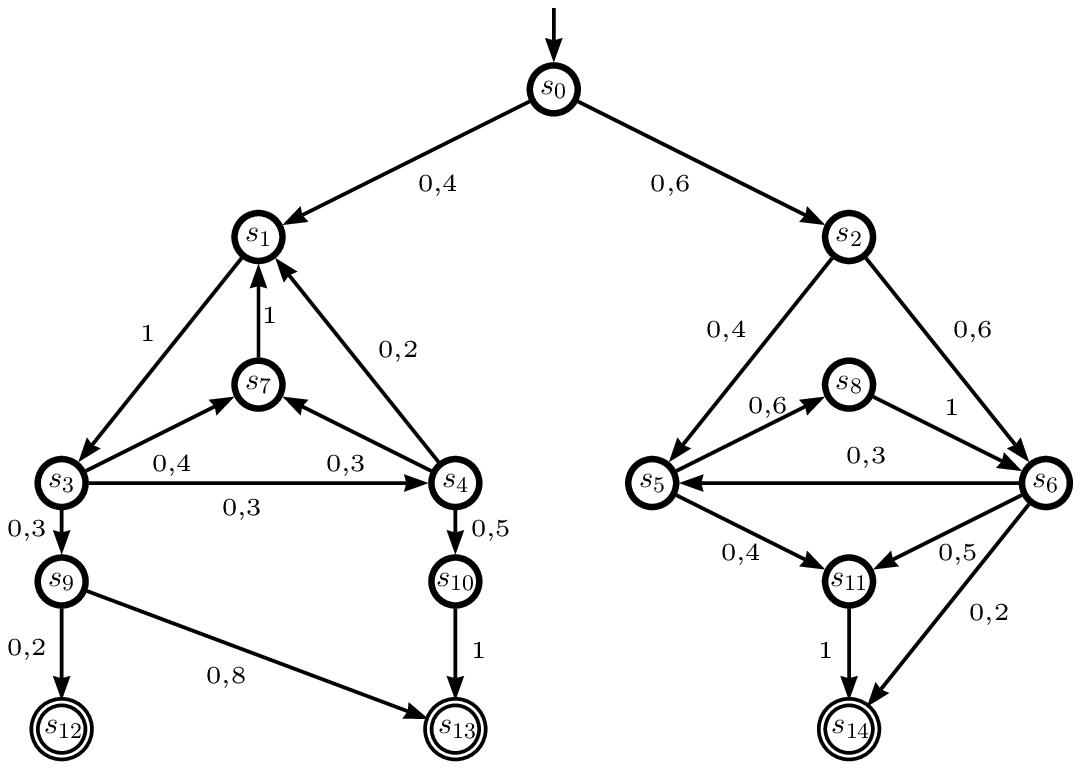}}
  \subfigure[Derived Acyclic $\MC$]{\label{fig:red-b}\includegraphics[width=6cm]{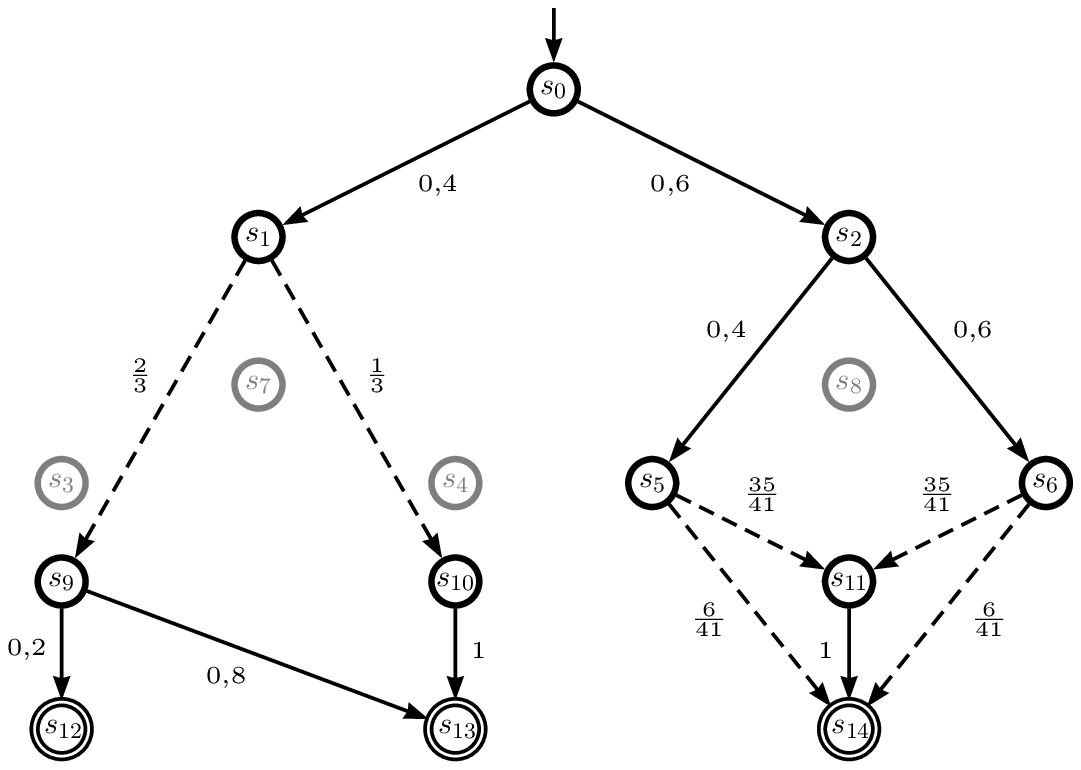}}
  \label{fig:reducingMCs}
  \caption{}
  \vspace{-0.5cm}
\end{figure}

\begin{example} Consider the $\MC$ $\mc$ of Figure \ref{fig:red-a}. The strongly connected components of $\mc$ are $\scc_1\eqdef\lbrace
s_1,s_3,s_4,s_7\rbrace$, $\scc_2\eqdef\lbrace s_5,s_6,s_8\rbrace$
and the singletons $\lbrace s_0 \rbrace$, $\lbrace s_2 \rbrace$,
$\lbrace s_9 \rbrace$, $\lbrace s_{10} \rbrace$, $\lbrace s_{11}
\rbrace$, $\lbrace s_{12} \rbrace$, $\lbrace s_{13} \rbrace$, and
$\lbrace s_{14} \rbrace$. The input states of $\scc_1$ are
$\Inp_{\scc_1}=\lbrace s_1\rbrace$ and its output states are
$\Out_{\scc_1}=\lbrace s_9,s_{10}\rbrace$. For $\scc_2$,
$\Inp_{\scc_2}=\lbrace s_5, s_6\rbrace$ and $\Out_{\scc_2}=\lbrace
s_{11},s_{14}\rbrace$. The reduced acyclic $\MC$ of $\mc$ is shown
in Figure \ref{fig:red-b}.
\end{example}

\subsection*{Rails and Torrents}

We now relate (finite) paths in $\Acyclic{\mc}$ (rails) to sets of
(finite) paths in $\mc$ (torrents).

\begin{dfn}[Rails] Let $\mc$ be a $\MC$. A finite path $\sigma\in\FPaths{\Acyclic{\mc}}$ will be called
a \emph{rail of }$\mc$.
\end{dfn}

Consider a rail $\sigma$, i.e., a finite path of $\Acyclic{\mc}$.
We will use $\sigma$ to represent those paths $\omega$ of $\mc$
that behave ``similar to'' $\sigma$ outside $\SCCs$ of $\mc$.
Naively, this means that $\sigma$ is a subsequence of $\omega$.
There are two technical subtleties to deal with: every input state
in $\sigma$ must be the first state in its $\SCC$ in $\omega$
(freshness) and every $\SCC$ visited by $\omega$ must be also
visited by $\sigma$ (inertia) (see Definition~\ref{dfn:sim}). We
need these extra conditions to make sure that no path $\omega$
behaves ``similar to'' two distinct rails (see
Lemma~\ref{lem:disjoint}).

Recall that given a finite sequence $\sigma$ and a (possible
infinite) sequence $\omega$, we say that $\sigma$ is a
\emph{subsequence} of $\omega$, denoted by $\sigma \subS \omega$,
if and only if there exists a strictly increasing function
$f:\lbrace 0,1,\ldots,|\sigma|-1\rbrace\rightarrow \lbrace
0,1,\ldots,|\omega|-1\rbrace$ such that $\forall_{0\leq i <
|\sigma|}. \sigma_i=\omega_{f(i)}$. If $\omega$ is an infinite
sequence, we interpret the codomain of $f$ as $\mathbb{N}$. In
case $f$ is such a function we write $\sigma \subS_f \omega$. Note
that finite paths and paths are sequences.

\begin{dfn} Let $\mc=(S,s_0,\M, L)$ be a $\MC$. On $S$ we consider the equivalence relation $\sim_{\!\!\!\!\!\!_{\mc}}$ satisfying $s\sim_{\!\!\!\!\!\!_{\mc}} t$ if and only if $s$ and $t$ are in the same
strongly connected component. Again, we usually omit the subscript
$\mc$ from the notation.
\end{dfn}


The following definition refines the notion of subsequence, taking
care of the two technical subtleties noted above.

\begin{dfn}\label{dfn:sim} Let $\mc=(S,s_0,\M,L)$ be a $\MC$, $\omega$ a (finite) path of $\mc$, and $\sigma\in \FPaths{\Acyclic{\mc}}$ a finite path of $\Acyclic{\mc}$. Then we write $\sigma\HsubS \omega$ if there exists $f:\lbrace 0,1,\ldots,|\sigma|-1\rbrace\rightarrow \mathbb{N}$ such that $\sigma\subS_f\omega$ and for all $0\leq i< |\sigma|$ we have
\begin{align*}
\forall_{0\leq j < f(i)}: \omega_{f(i)}\not\sim \omega_j & \mbox{; for all }i=0,1,\ldots|\sigma|-1, & [\mbox{\emph{Freshness property}}]\\
\forall_{f(i)< j< f(i+1)}:\omega_{f(i)}\sim \omega_{j} & \mbox{;
for all }i=0,1,\ldots|\sigma|-2. & [\mbox{\emph{Inertia
property}}]
\end{align*}
\noindent In case $f$ is such a function we write $\sigma\HsubS_f
\omega$.
\end{dfn}

\begin{example} Let $\mc=(S,s_0,\M,L)$ be the $\MC$ of Figure
\ref{fig:red-a} and take $\sigma=s_0 s_2 s_6 s_{14}$. Then for all
$i\in\mathbb{N}$ we have $\sigma\HsubS_{f_i} \omega_i$ where
$\omega_i=s_0 s_2 s_6 (s_5 s_8 s_6)^i s_{14}$ and $f_i(0)
\triangleq 0$, $f_i(1) \triangleq 1$, $f_i(2) \triangleq 2$, and
$f_i(3) \triangleq 3+3 i$. Additionally, $\sigma \not\HsubS s_0
s_2 s_5 s_8 s_6 s_{14}$ since for all $f$ satisfying
$\sigma\subS_f s_0 s_2 s_5 s_8 s_6 s_{14}$ we must have $f(2)=5$;
this implies that $f$ does not satisfy the freshness property.
Finally, note that $\sigma \not\HsubS s_0 s_2 s_6 s_{11} s_{14}$
since for all $f$ satisfying $\sigma\subS_f s_0 s_2 s_6 s_{11}
s_{14}$ we must have $f(2)=2$; this implies that $f$ does not
satisfy the inertia property.
\end{example}

We now give the formal definition of torrents.

\begin{dfn}[Torrents] Let $\mc=(S,s_0,\M, L)$ be a $\MC$ and $\sigma$ a sequence of states in $S$. We define the function
$\text{Torr}$ by
$$\Torrent{\mc,\sigma}\triangleq\lbrace \omega\in\Paths{\mc}\mid \sigma\HsubS \omega\rbrace.$$
\noindent We call $\Torrent{\mc,\sigma}$ the \emph{torrent}
associated to $\sigma$.
\end{dfn}

We now show that torrents are disjoint (Lemma~\ref{lem:disjoint}) and that
the probability of a rail is equal to the probability of its associated torrent (Theorem~\ref{thm:railSameProb}). For this last result, we first show that
torrents can be represented as the disjoint union of cones of finite paths. We call these finite paths \emph{generators} of the torrent (Definition~\ref{dfn:generators}).

\begin{lem}\label{lem:disjoint} Let $\mc$ be a $\MC$. For every $\sigma,\rho\in
\FPaths{\Acyclic{\mc}}$ we have
$$\sigma\not=\rho \Rightarrow\Torrent{\mc,\sigma}\cap \Torrent{\mc,\rho}=\emptyset$$
\end{lem}


\begin{dfn}[Torrent Generators]\label{dfn:generators} Let $\mc$ be a $\MC$. Then we define for every rail $\sigma\in
\FPaths{\Acyclic{\mc}}$ the set
$$\gen{\mc,\sigma}\triangleq\lbrace \rho\in \FPaths{\mc}\mid
\exists f: \sigma \HsubS_f \rho \land
f(|\sigma|-1)=|\rho|-1\rbrace.$$
\end{dfn}

In the example from the Introduction (see Figure~\ref{fig:MC related to MDP}), $s_0 s_1 s_3$ and $s_0 s_2 s_4$ are rails. The associated torrents are, respectively, $\{ s_0 s_1^n s_3^\omega \;|\; n \in \Nat^*\}$ and $\{ s_0 s_2^n s_4^\omega \;|\; n \in \Nat^*\}$ (note that $s_3$ and $s_4$ are absorbing states), i.e.~the paths going left and the paths going right. The generators of the first torrent are $\{ s_0 s_1^n s_3 \;|\; n \in \Nat^*\}$ and similarly for the second torrent.


\begin{lem}\label{lem:generators} Let $\mc$ be a $\MC$ and $\sigma\in \FPaths{\Acyclic{\mc}}$ a rail of $\mc$.
Then we have
$$\Torrent{\mc,\sigma}=\biguplus_{\rho\in\gen{\mc,\sigma}}\cyl{\rho}.$$
\end{lem}

\begin{thm}\label{thm:railSameProb} Let $\mc$ be a $\MC$. Then for every rail $\sigma\in\FPaths{\Acyclic{\mc}}$ we have
$$\measure{}{_{\Acyclic{\mc}}}{\cyl{\sigma}}=\measure{}{_{\mc}}{\Torrent{\mc,\sigma}}.$$
\end{thm}

\section{Significant Diagnostic Counterexamples}\label{sec:significantdiagnosticcounterexamples}

So far we have formalized the notion of paths behaving similarly
(i.e., behaving the same outside $\SCCs$) in a $\MC$ $\mc$ by
removing all $\SCC$ of $\mc$, obtaining $\Acyclic{\mc}$. A
representative counterexample to $\sat{\Acyclic{\mc}}{\leq
p}{\F\psi}$ will give rise to a representative counterexample to
$\sat{\mc}{\leq p}{\F\psi}$. For every finite path $\sigma$ in the
counterexample to $\sat{\Acyclic{\mc}}{\leq p}{\F\psi}$, the set
$\gen{\mc,\sigma}$ will be a witness. The union of these is the
representative counterexample to $\sat{\mc}{\leq p}{\F\psi}$.

Before giving a formal definition, there is still one technical
issue to resolve: we need to be sure that by removing $\SCCs$ we are not
discarding useful information. Because torrents are built from
rails, we need to make sure that when we discard $\SCCs$, we do
not discard rails that reach $\psi$.

We achieve this by first making states satisfying $\psi$
absorbing. Additionally, we make absorbing states from which it is
not possible to reach $\psi$. Note that this does not affect
counterexamples.

\begin{dfn}\label{dfn:abs mc}
Let $\mc=(S,s_0,\M,L)$ be a $\MC$ and $\psi$ a
propositional formula. We define the $\MC$
$\mc_\psi\triangleq(S,s_0,\M_\psi,L)$, with
$$ \M_\psi(s,t) \triangleq \left\lbrace
                           \begin{array}{ll}
                             1 & \mbox{if } s\not\in \SatF{\psi} \land s=t,\\
                             1 & \mbox{if } s\in \LangSt{\psi} \land s=t,\\
                             \M(s,t) & \mbox{if } s\in \SatF{\psi}-\LangSt{\psi},\\
                             0 & \mbox{otherwise,}\\
                             \end{array}
              \right. $$
\noindent where $\SatF{\psi}\eqdef \lbrace \allowbreak s\in S\mid
\allowbreak \allowbreak \measure{}{s}{\Reach{\mc,s, \allowbreak
\LangSt{\psi}}} \allowbreak >0\rbrace$ is the set of states
reaching $\psi$ in $\mc$.
\end{dfn}

The following theorem shows the relation between paths, finite
paths, and probabilities of $\mc$, $\mc_\psi$, and
$\Acyclic{\mc_\psi}$. Most importantly, the probability of a rail
$\sigma$ (in $\Acyclic{\mc_\psi}$) is equal to the probability of
its associated torrent (in $\mc$) (item \ref{i:5} below) and the
probability of $\F\psi$ is not affected by reducing $\mc$ to
$\Acyclic{\mc_\psi}$ (item \ref{i:6} below).

Note that a rail $\sigma$ is always a finite path in
$\Acyclic{\mc_\psi}$, but that we can talk about its associated
torrent $\Torrent{\mc_\psi,\sigma}$ in $\mc_\psi$ and about its
associated torrent $\Torrent{\mc,\sigma}$ in $\mc$. The former
exists for technical convenience; it is the latter that we are
ultimately interested in. The following theorem also shows that
for our purposes, viz.~the definition of the generators of the
torrent and the probability of the torrent, there is no difference
(items~\ref{i:3} and~\ref{i:4} below).

\begin{thm}\label{thm:pathInAbsSameProb} Let $\mc=(S,s_0,\M,L)$ be a $\MC$ and $\psi$ a propositional formula. Then for every
$\sigma\in\FPaths{\mc_\psi}$
\begin{enumerate}
\item $\FReach{\mc_\psi,s_0,\LangSt{\psi}}=\FReach{\mc,s_0,\LangSt{\psi}}$,
\item $\measure{}{_{\mc_\psi}}{\cyl{\sigma}}=\measure{}{_{\mc}}{\cyl{\sigma}}$,
\item\label{i:3} $\gen{\mc_\psi,\sigma}=\gen{\mc,\sigma}$,
\item\label{i:4} $\measure{}{_{\mc_\psi}}{\Torrent{\mc_\psi,\sigma}}=\measure{}{_{\mc}}{\Torrent{\mc,\sigma}}$,
\item\label{i:5} $\measure{}{\!_{\Acyclic{\mc_{\psi}}}\!}{\cyl{\sigma}}=\measure{}{_{\mc}}{\Torrent{\mc,\sigma}}$,
\item\label{i:6} $\sat{\Acyclic{\mc_{\psi}}}{\leq p}{\F \psi}$ if and only if $\sat{\mc}{\leq p}{\F \psi}$, for any $p\in[0,1]$.
\end{enumerate}
\end{thm}


\begin{dfn}[Torrent-Counterexamples]\label{dfn:torrentCounterexample} Let $\mc=(S,s_0,\M,L)$ be a $\MC$, $\psi$ a propositional formula, and $p\in[0,1]$ such that $\nonsat{\mc}{\leq p}{\F \psi}$.
Let $\ce$ be a representative counterexample to
$\sat{\Acyclic{\mc_{\psi}}}{\leq p}{\F \psi}$. We define the set
\begin{align*}
\TorRepCount{\ce}\eqdef \lbrace \gen{\mc,\sigma}\mid\sigma\in
\ce\rbrace.
\end{align*}
\noindent We call the set $\TorRepCount{\ce}$ a
\emph{torrent-counterexample} of $\ce$. Note that this set is a
partition of a counterexample to $\sat{\mc}{\leq p}{\F \psi}$.
Additionally, we denote by $\TorCountSet{\mc,p,\psi}$ to the set
of all torrent-counterexamples to $\sat{\mc}{\leq p}{\F \psi}$,
i.e., $\lbrace \TorRepCount{\ce}\mid \ce\in
\CountSet{\mc,p,\psi}\rbrace$.
\end{dfn}

\begin{thm}\label{thm:torrentCounterexample} Let $\mc=(S,s_0,\M,L)$ be a $\MC$, $\psi$ a propositional formula, and $p\in[0,1]$ such that $\nonsat{\mc}{\leq p}{\F \psi}$.
Take $\ce$ a representative counterexample to
$\sat{\Acyclic{\mc_{\psi}}}{\leq p}{\F \psi}$. Then the set of
finite paths $\biguplus_{W\in\TorRepCount{\ce}} W$ is a
representative counterexample to $\sat{\mc}{\leq p}{\F \psi}$.
\end{thm}

Note that for each $\sigma\in \ce$ we get a witness
$\gen{\mc,\sigma}$. Also note that the number of rails is finite, so
there are also only finitely many witnesses.

Following \cite{hk_2007_counterexamples}, we extend the notions of
\emph{minimum counterexamples}, \emph{strongest evidence} and
\emph{smallest counterexample} to torrents.

\begin{dfn}[Minimum torrent-counterexample] Let $\mc$ be a $\MC$,
$\psi$ a propositional formula and $p\in[0,1]$. We say that
$\ce_t\in\TorCountSet{\mc,p,\psi}$ is a \emph{minimum
torrent-counterexample} if $|\ce_t|\leq|\ce_t^\prime|$, for all
$\ce^\prime_t\in\TorCountSet{\mc,p,\psi}$.
\end{dfn}

\begin{dfn}[Strongest torrent-evidence] Let $\mc$ be a $\MC$,
$\psi$ a propositional formula and $p\in[0,1]$. A \emph{strongest
torrent-evidence} to $\smash{\nonsat{\mc}{\leq p}{\F \psi}}$ is a torrent
$W_\sigma\in\Torrent{\mc,\LangSt{\psi}}$ such that
$\measure{}{\mc}{W_\sigma}\geq \measure{}{\mc}{W_\rho}$ for all
$W_\rho \in \Torrent{\mc,\LangSt{\psi}}$.
\end{dfn}

Now we define our notion of significant diagnostic counterexamples.
It is the generalization of most indicative counterexample from
\cite{hk_2007_counterexamples} to our setting.

\begin{dfn}[Most indicative torrent-counterexample] Let $\mc$ be a $\MC$, $\psi$ a propositional formula and $p\in[0,1]$.
We call $\ce_t\in\TorCountSet{\mc,p,\psi}$ a \emph{most indicative
torrent-counterexample} if it is a minimum torrent-counterexample
and $\measure{}{}{\bigcup_{W\in\ce_t}\cyl{W}}\geq
\measure{}{}{\bigcup_{W\in\ce^\prime_t}\cyl{W}}$ for all minimum
torrent counterexamples $\ce^\prime_t\in\TorCountSet{\mc,p,\psi}$.
\end{dfn}


By Theorem \ref{thm:torrentCounterexample} it is possible to
obtain strongest torrent-evidence and most indicative
torrent-counterexamples of a $\MC$ $\mc$ by obtaining strongest
evidence and most indicative counterexamples of
$\Acyclic{\mc_\psi}$ respectively.

\section{Computing Counterexamples}
\label{sec:coumputing-counterexamples}

In this section we show how to compute most indicative torrent-counterexamples.
We also discuss what information to present to the user: how to present
witnesses and how to deal with overly large strongly connected components.

\subsection{Maximizing Schedulers}
\label{sec:max-schedulers}

The calculation of a maximal probability on a reachability problem
can be performed by solving a linear minimization
problem~\cite{ba_1995_probabilistic,dealfaro_1997_thesis}. This
minimization problem is defined on a system of inequalities that
has a variable $x_i$ for each different state $s_i$ and an
inequality $\sum_j\pi(s_j)\cdot x_j \leq x_i$ for each
distribution $\pi\in\tau(s_i)$.
The maximizing (deterministic memoryless) scheduler $\eta$ can be
easily extracted out of such system of inequalities after
obtaining the solution. If $p_0, \dots, p_n$ are the values that
minimize $\sum_ix_i$ in the previous system, then $\eta$ is such
that, for all $s_i$, $\eta(s_i)=\pi$ whenever $\sum_j\pi(s_j)\cdot
p_j = p_i$. In the following we denote $\Prob{}{s_i}{\F
\psi}\eqdef x_i$.
%




\subsection{Computing most indicative torrent-counterexamples}
\label{sec:computing}

We divide the computation of most indicative
torrent-counterexamples to $\sat{\mdp}{\leq p}{\F \psi}$ in three
stages: \emph{pre-processing}, \emph{$\SCC$ analysis}, and
\emph{searching}.

\paragraph{Pre-processing stage.}

We first modify the original $\MC$ $\mc$ by making all states in
$\LangSt{\psi} \cup S \setminus \SatF{\psi}$ absorbing.  In this
way we obtain the $\MC$ $\mc_{\psi}$ from Definition~\ref{dfn:abs
mc}.  Note that we do not have to spend additional computational
resources to compute this set, since $\SatF{\psi} = \lbrace s\in
S\mid \Prob{}{s}{\psi} > 0 \rbrace$ and hence all required data is
already available from the $\LTL$ model checking phase.

\paragraph{$\SCC$ analysis stage.}

We remove all $\SCCs$ $\scc$ of $\mc_\psi$ keeping just
\emph{input states} of $\scc$, getting the acyclic $\MC$
$\Acyclic{\mc_\psi}$ according to Definition~\ref{dfn:acyclicMC}.

To compute this, we first need to find the $\SCCs$ of $\mc_\psi$.
There exists well known algorithms to achieve this: Kosaraju's,
Tarjan's, Gabow's algorithms (among others). We also have to
compute the reachability probability from input states to output
states of every $\SCC$. This can be done by using steady state
analysis techniques~\cite{cassandras_1993_steadystateanalysis}.



\paragraph{Searching stage.}

To find most indicative torrent-counterexamples in $\mc$, we find
most indicative counterexamples in $\Acyclic{\mc_\psi}$.  For this
we use the same approach as \cite{hk_2007_counterexamples},
turning the MC into a weighted digraph to exchange the problem of
finding the finite path with highest probability by a shortest
path problem.  The nodes of the digraph are the states of the
$\MC$ and there is an edge between $s$ and $t$ if $\M(s,t) > 0$.
The weight of such an edge is $-\log \M(s,t)$.

Finding the most indicative counterexample in $\Acyclic{\mc_\psi}$
is now reduced to finding $k$ shortest paths. As explained in
\cite{hk_2007_counterexamples}, our algorithm has to compute $k$
on the fly. Eppstein's algorithm \cite{epps_98_k-shortest-paths}
produces the $k$ shortest paths in general in $O(m + n \log n +
k)$, where $m$ is the number of nodes and $n$ the number of edges.
In our case, since $\Acyclic{\mc_\psi}$ is acyclic, the complexity
decreases to $O(m + k)$.

\subsection{Debugging issues}\label{sec:debugging}

\paragraph{Representative finite paths.}

What we have computed so far is a most indicative counterexample
to $\sat{\Acyclic{\mc_\psi}}{\leq p}{\F \psi}$. This is a finite
set of rails, i.e., a finite set of paths in $\Acyclic{\mc_\psi}$.
Each of these paths $\sigma$ represents a witness
$\gen{\mc,\sigma}$. Note that this witness itself has usually
infinitely many elements.


In practice, one somehow has to display a witness to the user. The
obvious way would be to show the user the rail $\sigma$. This,
however, may be confusing to the user as $\sigma$ is not a finite
path of the original Markov Decision Process. Instead of
presenting the user with $\sigma$, we therefore show the user the
element of $\gen{\mc,\sigma}$ with highest probability.

\begin{dfn} Let $\mc$ be a $\MC$, and $\sigma\in\FPaths{\Acyclic{\mc_\psi}}$ a rail of $\mc$.
We define the \emph{representant of} $\Torrent{\mc,\sigma}$ as
\begin{align*}
\repTorrent{\mc,\sigma}=\repTorrent{\biguplus_{\rho\in\gen{\mc,\sigma}}\cyl{\rho}}\eqdef
\arg \max_{\rho\in\gen{\mc,\sigma}}\measure{}{}{\cyl{\rho}}
\end{align*}
\end{dfn}

Note that given $\repTorrent{\mc,\sigma}$, one can easily recover
$\sigma$. Therefore, no information is lost by presenting torrents
as a single element of the torrent instead of as a rail.

\paragraph{Expanding $\SCC$.}

\begin{wrapfigure}{r}{4cm}
\vspace{-1cm}
\includegraphics[width=4cm]{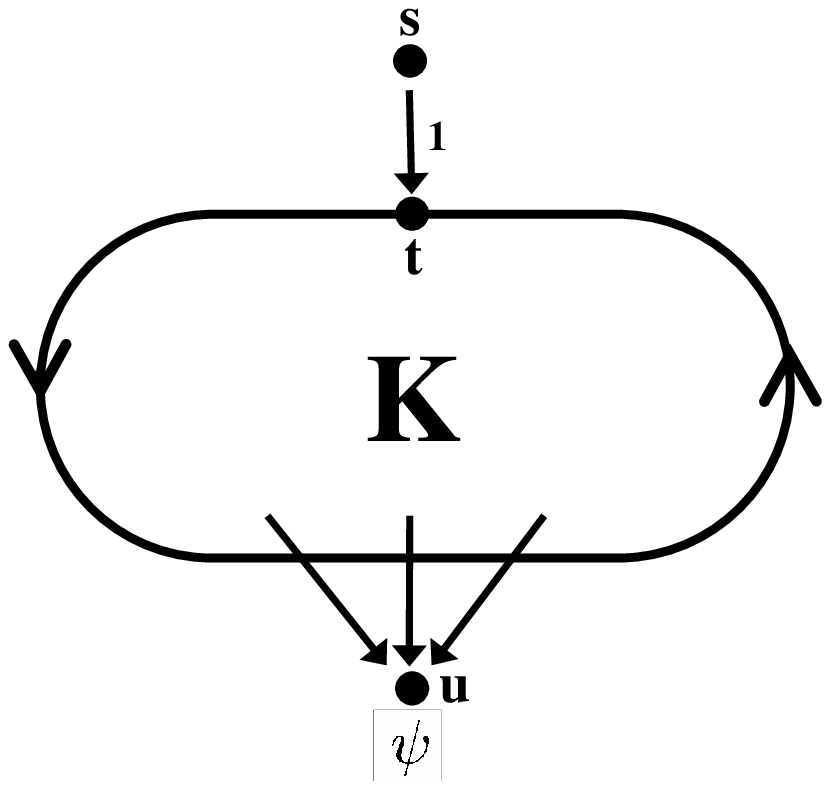}
\caption{} \label{fig:bigSCC} \vspace{-0.5cm}
\end{wrapfigure}

It is possible that the system contains some very large strongly
connected components. In that case, a single witness could have a
very large probability mass and one could argue that the
information presented to the user is not detailed enough. For
instance, consider the Markov chain of Figure~\ref{fig:bigSCC}
in which there is a single large $\SCC$ with input state $t$ and
output state $u$.

The most-indicative torrent counterexample to the property
$\sat{\mc}{\leq 0.9}{\F \psi}$ is simply $\{\gen{s t u}\}$, i.e.,
a single witness with probability mass 1 associated to the rail $s
t u$. Although this may seem uninformative, we argue that it is
more informative than listing several paths of the form $s t
\cdots u$ with probability summing up to, say, $0.91$. Our single
witness counterexample suggests that the outgoing edge to a state
not reaching $\psi$ was simply forgotten; the listing of paths
still allows the possibility that one of the probabilities in the
whole system is simply wrong.
\medskip

Nevertheless, if the user needs more information to tackle bugs
inside strongly connected components, note that there is more
information available at this point. In particular, for every
strongly connected component $\scc$, every input state $s$ of
$\scc$ (even for every state in $\scc$), and every output state
$t$ of $\scc$, the probability of reaching $t$ from $s$ is already
available from the computation of $\Acyclic{\mc_\psi}$ during the
$\SCC$ analysis stage of Section~\ref{sec:computing}.

\section{Final Discussion}
\label{sec:conclusions}

We have presented a novel technique for representing and computing
counterexamples for nondeterministic and probabilistic systems. We
partition a counterexample in witnesses and state five properties
that we believe good witnesses should satisfy in order to be
useful as debugging tool: (similarity) elements of a witness
should provide similar debugging information; (originality)
different witnesses should provide different debugging
information; (accuracy) witnesses with higher probability should
indicate system behavior more likely to contain errors;
(significance) probability of a witness should be relatively high;
(finiteness) there should be finitely many witnesses. We achieve
this by grouping finite paths in a counterexample together in a
witness if they behave the same outside the strongly connected
components.

Presently, some work has been done on counterexample generation
techniques for different variants of probabilistic models
(Discrete Markov chains and Continues Markov chains)
\cite{ahl_2005_counterexamples,al_2006_search,hk_2007_counterexamples,hk_2007_counterexamplesDTMC}.
In our terminology, these works consider witnesses consisting of a
\emph{single} finite path.
We have already discussed in the Introduction that the single path
approach does not meet the properties of accuracy, originality,
significance, and finiteness.

Instead, our witness/torrent approach provides a high level of
abstraction of a counterexample.
By grouping together finite paths that behave the same outside
strongly connected components in a single witness, we can achieve
these properties to a higher extent.
Behaving the same outside strongly connected components is a
reasonable way of formalizing the concept of providing
\emph{similar}
debugging information. %
This grouping also makes witnesses significantly different form
each other: each witness comes form a different rail and each rail
provides a different way to reach the undesired property. Then
each witness provides \emph{original} information. Of course, our
witnesses are more \emph{significant} than single finite paths,
because they are sets of finite paths.
This also gives us more \emph{accuracy} than the approach with
single finite paths, as a collection of finite paths behaving the
same and reaching an undesired condition with high probability is
more likely to show how the system reaches this condition than
just a single path.  Finally, because there is a finite number of
rails, there is also a \emph{finite} number of witnesses.

Another key difference of our work to previous ones is that our
technique allows us to generate counterexamples for probabilistic
systems \emph{with} nondeterminism.
However, a recent report \cite{al_2007_counterexamplesMDP} also
considers counterexample generation for $\MDPs$. This work is
limited to upper bounded $\pCTL$ formulas without nested temporal
operators.
Besides, their technique significantly differs from ours.

Finally, among the related work, we would like to stress the
result of \cite{hk_2007_counterexamples}, which provides a
systematic characterization of counterexample generation in terms
of shortest paths problems. We use this result to generate
counterexamples for the acyclic Markov Chains.



In the future we intend to implement a tool to generate our
significant diagnostic counterexamples; a very preliminary version
has already been implemented. There is still work to be done on
improving the visualization of the witnesses, in particular, when
a witness captures a large strongly connected component.
Another direction is to investigate how this work can be extended
to timed systems, either modeled with continuous time Markov
chains or with probabilistic timed automata.

\bibliographystyle{alpha}
\bibliography{Counterexamples}

%
%
\newpage
\section*{Appendix: Proofs}

In this appendix we give proofs of the results that were omitted
from the paper for space reasons.

\begin{obs}\label{obs:scc-rails}
Let $\mc$ be a $\MC$. Since $\Acyclic{\mc}$ is acyclic we have
$\sigma_i \not \sim \sigma_j$ for every $\sigma\in
\FPaths{\Acyclic{\mc}}$ and $i\not=j$ (with the exception of absorbing
states).
\end{obs}

\begin{obs}\label{obs:surSCC} Let $\sigma,\omega$ and $f$ be such that $\sigma\HsubS_f
\omega$. Then $\forall i: \exists j: \omega_i\sim \sigma_j$. This
follows from $\sigma\subS_f \omega$  and the inertia property.
\end{obs}

\begin{lem} \label{lem:torrents}Let $\mc$ be a $\MC$, and $\sigma t s\in\FPaths{\Acyclic{\mc}}$. Additionally
let $\Delta_{\sigma t s}\triangleq\lbrace
\rho\tail{\pi}|\rho\in\gen{\sigma
t},\pi\in\FPaths{\SCC^+_t,t,\lbrace s\rbrace}\rbrace$. Then
$\Delta_{\sigma t s}=\gen{\sigma t s}$.
\end{lem}

\begin{proof}.

\noindent $(\ \supseteq\ )$ Let $\rho_0 \rho_1 \cdots \rho_k \in
\gen{\sigma t s}$ and $n_t$ the lowest subindex of $\rho$ such
that $\rho_{n_t}=t$. Take $\rho\triangleq\rho_0 \rho_1 \cdots
\rho_{n_t}$ and $\pi\triangleq\rho_{n_t} \cdots \rho_k$ (Note that
$\rho_0 \rho_1 \cdots \rho_k=\rho\tail{\pi}$). In order to prove
that $\rho_0 \rho_1 \cdots \rho_k \in \Delta_{\sigma t s}$ we need
to prove that

\begin{enumerate}
\item[(1)]{$\rho\in \gen{\sigma t}$, and}
\item[(2)]{$\pi\in \FPaths{\SCC^+_t,t,\lbrace s\rbrace}$.}
\end{enumerate}

    \begin{itemize}
    \item [(1)] Let $f$ be such that $\sigma t s \HsubS_f \rho_0 \rho_1 \cdots \rho_k$ and $f(|\sigma t s| -1)=k$. Take $g:\lbrace
    0,1,\ldots,|\sigma t|-1\rbrace\rightarrow \mathbb{N}$ be the restriction of $f$ . It is easy to check that $\sigma t \HsubS_g
    \rho$.
    Additionally $f(|\sigma t|-1)=n_t$ (otherwise
    $f$ would not satisfy the freshness property for $i=|\sigma
    t|-1$). Then, by definition of $g$, we have $g(|\sigma t|-1)=n_t$.

    \item [(2)] It is clear that $\pi$ is a path from $t$ to $s$.
    Therefore we only have to show that every state of $\pi$ is in
    $\SCC^+_t$. By definition of $\SCC^+_t$, $\pi_0=t\in\SCC^+_t$ and
    $s\in\SCC^+_t$ since $s\in\Out_{\SCC^+_t}$.
    Additionally, since $f$ satisfies inertia property we have
    that $\forall_{f(|\sigma t|-1)<j<f(|\sigma t s|-1)}:\rho_{f(|\sigma t|-1)}\sim
    \rho_j$, since $f(|\sigma t|-1)=n_t$ and
    $\pi\triangleq\rho_{n_t}\cdots\rho_k$ we have $\forall_{0<j<|\pi|-1}:t\sim
    \pi_j$ proving that $\pi_j\in\SCC^+_t$ for
    $j\in\lbrace1,\cdots,|\pi|-2\rbrace$.

    \comment{Alternative prove: Suppose that there exists $i$ such that $\pi_i\not \in
    \SCC^+_t$. But then $\rho_{f(n_t)}\not\sim \rho_{f(n_t)+i}$
    contradicting the Inertia property of $f$.}

    \end{itemize}

\noindent $(\ \subseteq \ )$ Take $\rho\in \gen{\sigma t}$ and
$\tail{\pi}\in \FPaths{\SCC^+_t,t,\lbrace s\rbrace}$. In order to
prove that $\rho\tail{\pi}\in \gen{\sigma t s}$ we need to show
that there exists a function $g$ such that:

\begin{enumerate}
\item[(1)]{$\sigma t s\HsubS_g\rho\tail{\pi}$,}
\item[(2)]{$g(|\sigma t s|-1)=|\rho\tail{\pi}|-1$.}
\end{enumerate}

Since $\rho\in \gen{\sigma t}$ we know that there exists $f$ be
such that $\sigma t \HsubS_f \rho$ and $f(|\sigma t|-1)=|\rho|-1$.
We define $g:\lbrace 0,1,\ldots,|\sigma t s|-1\rbrace\rightarrow
\lbrace 0,1,\ldots,|\rho\tail{\pi}|-1\rbrace$ by
     \begin{align*}
        g(i) & \eqdef \left\lbrace
                                       \begin{array}{ll}
                                         f(i) & \mbox{ if } i<|\sigma t s| - 1,\\
                                         |\rho\tail{\pi}|-1 & \mbox{ if }i=|\sigma t s| -1.\\
                                       \end{array}
                                   \right.
     \end{align*}

   \begin{itemize}

   \item [(1)] It is easy to check that $\sigma t s\subS_g \rho\tail{\pi}$. Now we will show that $g$
    satisfies Freshness and Inertia properties.

    \underline{Freshness property:} We need to show that for all $0\leq i<|\sigma t s|$ we have $\forall_{0\leq j<g(i)}: \rho\tail{\pi}_{g(i)}\not
    \sim\rho\tail{\pi}_j$. For the cases $i\in\lbrace
    0,\ldots,|\sigma t|-1\rbrace$ this holds since $\sigma t\HsubS_f
    \rho$ and definition of $g$.

    Consider $i=|\sigma t s|-1$, in this case we have to prove
    $\forall_{0\leq j< |\rho\tail{\pi}|-1}:
    \rho\tail{\pi}_{|\rho
    \tail{\pi}|-1)}\not \sim \rho\tail{\pi}_j$ or equivalently $\forall_{0\leq j< |\rho\tail{\pi}|-1}:
    s\not \sim \rho\tail{\pi}_j$.

       \begin{itemize}
       \item [Case $j\in\lbrace |\rho|,\ldots|\rho\tail{\pi}|-1\rbrace$]$\\$
        Since $\pi\in\FPaths{\SCC^+_t,t,\lbrace s\rbrace}$ and $s\in\Out^+_{\SCC^+_t}$ we have $\forall_{0\leq j<|\tail{\pi}|-1}:s\not\sim \tail{\pi}_j$

       \item [Case $j\in\lbrace 0,\ldots,|\rho|-1\rbrace$]$\\$
        Since $\sigma t s\in \FPaths{\Acyclic{\mc}}$ and Observation
        \ref{obs:scc-rails} we have $\forall_{0\leq j <|\sigma
        t|-1}:s\not\sim \sigma t_j$. Additionally, $\sigma t \HsubS_f
        \rho$, def. $g$, and Observation \ref{obs:surSCC} imply
        $\forall_{0\leq j <|\rho|}:s\not\sim \rho_j$ or equivalently
        $\forall_{0\leq j <|\rho|}:s\not\sim \rho\tail{\pi}_j$.

       \end{itemize}

    \comment{Alternative prove (by absurd) Suppose that there exists
    $j<g(|\sigma s|-1)$ such that
    $\rho\tail{\pi}_j\sim
    \rho\tail{\pi}_{g(|\sigma s|-1)}$. Since
    $g(|\sigma|-1)=|\rho|-1$ and $\pi\in
    \Paths{\SCC^+_t,t,\lbrace s\rbrace}$ we have that $j<|\rho|-1$.
    Additionally, note that since $\rho_j {\sigma}_{j+1}
    \cdots \rho_{|\rho|-1}$ is a finite
    path from $\rho_j$ to $t$ and $\pi$ is a finite path
    from $t$ to $s$ we have that $t\sim s$ which contradicts
    Observation \ref{obs:scc-rails}.}

    \underline{Inertia property:} Since $\pi\in\FPaths{SCC^+_t,t,\lbrace s\rbrace }$ we
    know that $\forall_{0\leq j < |\pi|-1}: t\sim \pi_j$ which implies
    that
    $\forall_{|\rho|-1<j<|\rho\tail{\pi}|-1}:
    \rho\tail{\pi}_{|\rho|-1}\sim
    \rho\tail{\pi}_j$ or equivalently
    $\forall_{g(|\sigma|-1)<\allowbreak j<\allowbreak
    g(|\sigma s|-1)}:
    \rho\tail{\pi}_{g(|\rho|-1)}\sim
    \rho\tail{\pi}_j$ showing that $g$ satisfies
    the inertia property.

   \item [(2)] Follows from the definition of $g$.

    \end{itemize}

\end{proof}

\noindent{\bf Theorem \ref{thm:railSameProb}.~}{\em Let
$\mc=(S,s_0,\M,L)$ be a $\MC$. Then for every rail
$\sigma\in\FPaths{\Acyclic{\mc}}$ we have
$$\measure{}{_{\Acyclic{\mc}}}{\cyl{\sigma}}=\measure{}{_{\mc}}{\Torrent{\sigma}}.$$

\begin{proof} By induction on the structure of $\sigma$.
\begin{itemize}

\item [\emph{Base Case:}]

$\measure{}{\Acyclic{\mc}}{\cyl{s_0}}=\measure{}{\Acyclic{\mc}}{\Paths{\Acyclic{\mc},s_0}}=1=\measure{}{\mc}{\Paths{\mc,s_0}}=\measure{}{\mc}{\Torrent{s_0}}.$

\item [\emph{Inductive Step:}]

Let $t$ be such that $\last{\sigma}=t$.
Suppose that $t\in S_{\Com}$. Then
$$\begin{array}{rclr}
\hbox to 0pt{$\measure{}{\Acyclic{\mc}}{\cyl{\sigma s}}$\hss} \\
&=&\measure{}{\Acyclic{\mc}}{\cyl{\sigma}}\cdot \Acyclic{\M}(t,s)\\
&=&\measure{}{\mc}{\Torrent{\sigma}}\cdot \M(t,s)&\\
&&&\llap{$\{\text{Inductive Hypothesis and definition of }\M\}$}\\
&=&\measure{}{\mc}{\biguplus_{\rho\in\gen{\sigma}} \cyl{\rho}}\cdot \M(t,s)&\explan{\text{Lem.~}\ref{lem:generators}}\\
&=&\sum_{\rho\in\gen{\sigma}}\measure{}{\mc}{\cyl{\rho}}\cdot \measure{}{\mc}{\cyl{ts}}\\
&=&\sum_{\rho\in\gen{\sigma}}\measure{}{\mc}{\cyl{\rho \tail{ts}}}&\\
&&&\llap{$\{\text{Distributivity and }\last{\rho}=t \text{ for all }\rho\in\gen{\sigma}\}$}\\
&=&\sum_{\rho\in\gen{\sigma},\pi\in\Paths{\SCC^+_t,t,\lbrace s\rbrace}}\measure{}{\mc}{\cyl{\rho \tail{\pi}}}\\
&=&\sum_{\rho\in \Delta_{\sigma s}}\mu_{_{\mc}}(\cyl{\rho})&\explan{\text{Dfn.~}\Delta}\\
&=&\sum_{\rho\in \gen{\sigma s}}\mu_{_{\mc}}(\cyl{\rho})&\explan{\text{Lem.~}\ref{lem:torrents}}\\
&=&\measure{}{_{\mc}}{\biguplus_{\rho\in \gen{\sigma s}}\cyl{\rho}}\\
&=&\measure{}{_{\mc}}{\Torrent{\sigma s}}&\explan{\text{Lem.~}\ref{lem:generators}}\\
\end{array}$$
Now suppose that $t\in S_{\Inp}$. We denote by $\Acyclic{\M}$ to
the probability matrix of $\Acyclic{\mc}$, then
$$\begin{array}{rclr}
\hbox to 0pt{$\measure{}{_{\Acyclic{\mc}}}{\cyl{\sigma s}}$\hss} \\
&=&\measure{}{_{\Acyclic{\mc}}}{\cyl{\sigma}}\cdot \Acyclic{\M}(t,s)\\
&=&\measure{}{_{\mc}}{\Torrent{\sigma}}\cdot \Acyclic{\M}(t,s)&\explan{\text{HI}}\\
&=&\measure{}{_{\mc}}{\biguplus_{\rho\in\gen{\sigma}}\cyl{\rho}}\cdot\Acyclic{\M}(t,s)&\explan{\text{Lem.~\ref{lem:generators}}}\\
&=&\left(\sum_{\rho\in\gen{\sigma}}\measure{}{_{\mc}}{\cyl{\rho}}\right)\cdot\Acyclic{\M}(t,s)\\
&=&\sum_{\rho\in \gen{\sigma}}\measure{}{_{\mc}}{\cyl{\rho}}\cdot \measure{}{_{\mc,t}}{\Paths{\SCC^+_t,t,\lbrace s\rbrace}}&\\
&&&\llap{$\{\text{By definition of }\Acyclic{\M} \text{ and distributivity}\}$}\\
&=&\sum_{\rho\in \gen{\sigma}}\measure{}{_{\mc}}{\cyl{\rho}}\cdot \sum_{\pi\in \FPaths{\SCC^+_t,t,\lbrace s\rbrace}}\measure{}{_{\mc,t}}{\cyl{\pi}}\\
&=&\sum_{\rho\in \gen{\sigma},\pi\in\FPaths{\SCC^+_t,t,\lbrace s\rbrace}}\measure{}{_{\mc}}{\cyl{\rho \tail{\pi}}}&\explan{\text{Dfn.~} \mu}\\
&=&\sum_{\rho\in \Delta_{\sigma s}}\measure{}{_{\mc}}{\cyl{\rho}}&\explan{\text{Dfn.~}\Delta}\\
&=&\sum_{\rho\in \gen{\sigma s}}\measure{}{_{\mc}}{\cyl{\rho}}&\explan{\text{Lem.~\ref{lem:torrents}}}\\
&=&\measure{}{_{\mc}}{\biguplus_{\rho\in \gen{\sigma s}}\cyl{\rho}}\\
&=&\measure{}{_{\mc}}{\Torrent{\sigma s}}&\explan{\text{Lem.~\ref{lem:generators}}}\\
\end{array}$$
\end{itemize}

\end{proof}

\end{document}